\documentstyle[aps,pra,subfigure,epsfig,amsmath,amssymb]{revtex}

\newcommand {\be}{\begin{equation}}
\newcommand {\ee}{\end{equation}}
\newcommand {\bea}{\begin{eqnarray}}
\newcommand {\eea}{\end{eqnarray}}

\newcommand {\vett}[1] {{\mathbf{#1}}}

\newcommand {\Tr}{{\rm Tr}}
\renewcommand {\Im}{{\rm Im}} 
\renewcommand {\Re}{{\rm Re}} 

\draft

\begin{document}

\title{Light scattering from  a degenerate quasi-onedimensional 
  confined gas of non-interacting fermions}
\author{Patrizia Vignolo, Anna Minguzzi and M. P. Tosi}
\address{Istituto Nazionale di Fisica della Materia and Classe di Scienze,
Scuola Normale Superiore,
Piazza dei Cavalieri 7, I-56126 Pisa, Italy}
\maketitle

\begin{abstract}
We evaluate the scattering functions of a gas of spin-polarized,
non-interacting fermions confined in a quasi-onedimensional 
harmonic trap at zero temperature. The main  focus is on the inelastic
scattering spectrum and on the angular distribution of scattered light
from a mesoscopic atomic cloud 
as probes of its discrete quantum
levels and of its shell structure  in this restricted geometry. The
dynamic structure factor  
is calculated and compared with the results of a
local-density approximation exploiting the spectrum of  a one-dimensional
 homogeneous Fermi gas. The elastic and inelastic contributions to
the static structure factor are separately evaluated: the inelastic
term becomes dominant as the momentum transfer increases, masking
a small peak at twice the Fermi momentum in the elastic term but still
reflecting  the discrete quantum levels of the cloud.
A fast and accurate numerical method is
developed to evaluate the full pair distribution function for
mesoscopic inhomogeneous clouds.

\end{abstract}
\pacs{PACS numbers: 05.30.Fk,32.90.+a}

\section{Introduction}
There currently is considerable experimental activity in trapping
and cooling dilute gases of the fermionic alkali atoms $^{40}$K
\cite{jila_exp} and 
$^6$Li \cite{paris_exp} in the
quantum degeneracy regime. A quasi-onedimensional (quasi-1D) and almost
ideal neutral Fermi gas can be realized in these experiments by using
a very anisotropic, cigar-shaped magnetic trap. In this arrangement
the fermionic atoms can be placed into a single Zeeman sublevel,
yielding a spin-polarized state in which their mutual interactions are
extremely weak \cite{demarco_pra,stoof_varenna}, and are subject to
approximately harmonic confinement in the axial direction and to tight
confinement in the other directions. 

The possibility of experimentally studying such a close realization of
an ideal 1D Fermi gas in harmonic external potential presents great
interest for the theory of quantum gases. To begin with, the 1D model
of non-interacting fermions has the same particle density profile and
thermodynamic and dynamic properties as the 1D model of point-like
bosons with hard-core interactions
\cite{girardeau}.
Given this wider
relevance of the model, harmonic confinement is probably the simplest
way to realize an inhomogeneous quantum fluid on which to test the
ideas and the implementations  of Density Functional Theory
\cite{dft}. Of specific interest in the present context is the kinetic
energy functional, which can be constructed exactly from the particle
density in the case of $N$ fermions moving independently in a
harmonic oscillator potential \cite{lawes-march_2d}. Finally, the
shell structure that was noticed in the particle density of the Fermi
gas in 3D \cite{schneider-bruun} is greatly enhanced in lower
dimensionality \cite{noi,gleisberg}. Very efficient
numerical methods have been developed \cite{noi,papirone1} to evaluate
ground-state properties of mesoscopic 1D clouds containing large
numbers of fermions. 

The present work completes our earlier study of the ideal 1D Fermi gas
under harmonic confinement \cite{noi,papirone1} by evaluating the
spectral properties of low-frequency radiation scattered from its
quasi-1D realization.  Previous studies of the scattering spectrum of the
homogeneous ideal Fermi gas \cite{libro_pines,javanainen} and of the
off-resonant light scattering properties of an ideal Fermi gas under
isotropic harmonic confinement \cite{demarco_pra} have referred to the 3D 
 case.   Our objective is to examine how the shell
structure of the particle density profile 
and the single
occupancy of discrete quantum levels in this specific quasi-1D geometry 
are reflected in the
light-scattering properties and, at a deeper level, how the shell
structure appears in the pair distribution function as a gauge of the
spatial correlations induced by the Pauli exclusion principle between
pairs of fermions. 

The layout of the paper is briefly as follows. In Sections~\ref{sec2}
and~\ref{sec3} we recall from the work of Javanainen and Ruostekoski
\cite{javanainen} the basic facts about light scattering from an
atomic cloud and explicitly set out how the inelastic scattering
spectrum  and the angular distribution of the scattered light are
related to the time-dependent pair distribution function.
In Sec.~\ref{secsw} we present
an evaluation of the dynamic structure factor of a mesoscopic
 Fermi gas  in harmonic confinement 
and compare its results with those obtained from a
local-density approximation (LDA) exploiting the spectrum of the
homogeneous gas. Section~\ref{sk}  evaluates both the elastic and
the inelastic contribution to the static structure factor, again comparing
the latter results with the LDA. In Sec.~\ref{sec6} we present
numerical results for the  equal-times pair distribution function,
which are obtained by an extension of the Green's function method previously
developed \cite{noi} for the one-body density matrix. Finally,
Sec.~\ref{sec7} summarizes our main conclusions.

\section{Light scattering from an atomic cloud}
\label{sec2}
\begin{figure}
\centerline{\psfig{file=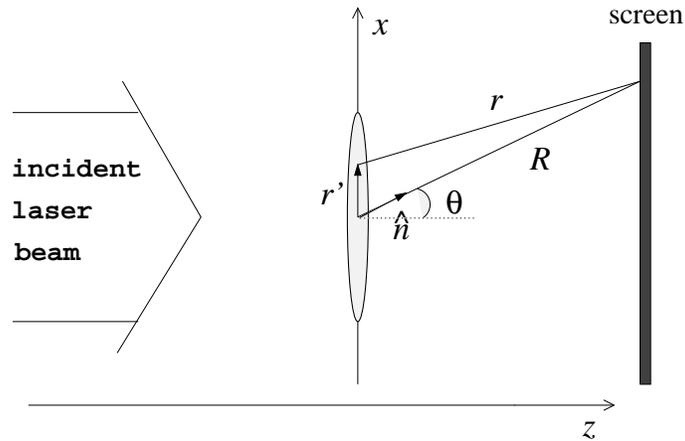,width=0.5\linewidth}}
\caption{Geometry of the light scattering configuration described in
the text.}
\label{fig_1}
\end{figure}

We consider a confined cloud of spin-polarized, magnetically trapped
atoms in a quasi-1D configuration.
We focus on a light scattering experiment with the 
incident beam  propagating along the $z$ axis,
 orthogonal to the $x$ axis which is the axial direction  of the
magnetic trap  
(see Fig.~\ref{fig_1}).
The positive-frequency component of the incident electric field is
\begin{equation}
\vett E_F^+(z,t)= \frac{1}{2} {\cal E} \hat \epsilon e^{i(k_Lz-\omega_Lt)}\;,
\end{equation}
where ${\cal E}$ is the field amplitude, $\hat \epsilon$ is the
polarization, $k_L$ and $\omega_L$ are the light wave number and
frequency. 
In regard to the internal states we model each atom as a
two-level system, {\it i.e.} 
an excited state $|e\rangle$ and a ground state $|g \rangle$ separated
by an energy $\hbar \omega_A$. This corresponds to a suitable
choice of the Zeeman sublevel for the atoms in the ground state and, at
the same time, of the light polarization.
Angular momentum degeneracy of the
internal states in the case of an unpolarized cloud has been treated
by Javanainen and Ruostekoski \cite{javanainen}. 

In the Born approximation the scattered light at a point on the screen
designated by the vector $\vett R$ (see again Fig.~\ref{fig_1}) is
given by \cite{javanainen}
\begin{equation}
\vett E_S^+(\vett R,t)={\cal E}\frac{k_L^2}{8 \pi \epsilon_0\hbar
\delta}\frac{e^{ik_LR}}{R}[\hat  n\times (\hat  n\times
{\mathbf d}^-)] \,(\hat {\mathbf \epsilon} \cdot {\mathbf
d}^+)\int_{{\rm \!cloud\!}}d^3r'\,e^{-i {\mathbf k} \cdot
{\mathbf  r}' }\hat \psi^\dagger(\vett r',t) \hat \psi(\vett r,t)\;.
\end{equation}
Here $\delta =\omega_L-\omega_A$ is the detuning between the laser
frequency and the atomic transition, $\hat n=\vett R/R$, $\vett
d^+= {\cal D}\,\hat \epsilon |e\rangle\langle g|$ and $\vett
d^-={\cal D}\,\hat \epsilon |g\rangle\langle e|$ are the raising and lowering
parts of the dipole-moment operator of amplitude ${\cal D}$, $\vett
k=k_L \hat n-\vett k_L$ is the  wave vector transfer and $\vett r'$
indicates an atom in the cloud (see Fig.~\ref{fig_1}). Finally, $\hat
\psi(\vett r,t)$ is the annihilation field operator for atoms in the
internal state $|g\rangle$.

The spectrum of scattered radiation from atoms in the ground state
at the point $\vett R$ is
\begin{equation}
{\cal I}(\vett k,\omega)=\frac{1}{2\pi} M(\hat \epsilon,\hat
n)I(R)\frac{\Omega^2}{\delta^2}\int dt\int d^3r_1\int
d^3r_2\,e^{i(\omega t-{\mathbf k} \cdot ({\mathbf r}_1-{\mathbf
r}_2))}\langle \hat \psi^\dagger(\vett r_1,t)\hat \psi(\vett
r_1,t)\hat\psi^\dagger (\vett r_2,0)\hat \psi (\vett r_2,0)\rangle\;.
\label{s_kw}
\end{equation}
Here $M(\hat \epsilon,\hat n)=(1+\cos^2(\theta))/2$  is the geometric
factor for detection at an angle $\theta$ in the case of
circularly polarized light; 
$I(R)={\cal D}^2\omega_L^2/(32 \pi^2\epsilon_0c^3R^2)$ is the total
intensity scattered by a single dipole, and $\Omega={\cal D}{\cal E}/(2\hbar)$
is the Rabi frequency. The triple integral on the RHS of
Eq.~(\ref{s_kw}) at $\omega \neq 0$ is the dynamic structure
factor $S(\vett k,\omega)$ for inelastic scattering.

The angular distribution of the scattered light is 
\begin{equation}
{\cal I}(\vett k)=\int d\omega \,{\cal I} (\vett k, \omega)\;,
\label{essek}
\end{equation}
the angular dependence being  implicit in the direction of
$\vett k$.
The distribution of scattered light can be decomposed into two
components: an elastic term, originating from the diffraction of the
light by a finite object of given density profile, and an inelastic
term determined by the excitations of the quantum fluid.

\section{General definitions}
\label{sec3}
According to  Eq.~(\ref{s_kw}) the microscopic quantity which
is relevant  for the spectrum of scattered light is the time-dependent
pair distribution function,
\begin{equation}
\rho_{2}(1,2)=\langle \hat \psi^\dagger (1) \hat \psi (1) \hat
\psi^\dagger (2) \hat \psi(2)\rangle\;.
\end{equation}
Here we have set $(1)=(\vett r_1,t_1)$ and $(2)=(\vett r_2,t_2)$. 
For a non-interacting Fermi gas the 
expansion of  the field operators into energy modes $\hat a_i$, that
is $\hat \psi(1)=\sum_i\phi_i(\vett r_1)\exp(-i\varepsilon_i t_1) \hat
a_i$, yields an explicit expression for the pair distribution function
in terms of the single-particle orbitals $\phi_i(\vett r)$: 
\begin{eqnarray}
\rho_{2}(1,2)&=&\sum_{i,j} \phi^*_i(\vett r_1)\phi_j(\vett
r_1)\phi^*_j(\vett r_2) \phi_i(\vett r_2)
e^{-i(\varepsilon_j-\varepsilon_i)(t_1-t_2)/\hbar} 
f(\varepsilon_i) [1-f(\varepsilon_j)] +n(1)n(2)
\nonumber \\
&\equiv& S(1,2)+n(1)n(2)\;,
\end{eqnarray}
where $n(1)$ is the equilibrium density profile and
$f(\varepsilon)$ is the Fermi distribution
function. 

The  dynamic structure factor is
given by
\begin{eqnarray}
S(\vett k,\omega)&=&\int dt \int d^3r_1 \int d^3 r_2 \,e^{i[\omega t
-{\mathbf k}\cdot ({\mathbf r}_1-{\mathbf r}_2)]}
\left. S(1,2)\right|_{t=t_1-t_2} \nonumber \\
&=&\sum_{i,j}\left|\int d^3 r \, e^{-i {\mathbf  k}\cdot {\mathbf  r}}\phi^*_i(\vett
r)\phi_j(\vett r)\right|^2 f(\varepsilon_i) 
[1-f(\varepsilon_j)]\,2 \pi
\delta(\omega-(\varepsilon_j-\varepsilon_i)/\hbar) \;.
\label{s_i_kw}
\end{eqnarray}
By integration over  frequency one  obtains the inelastic
contribution to the static structure factor:
\begin{eqnarray}
S_i(\vett k)&\equiv &\int \frac{d\omega}{2\pi} \,S(\vett k,\omega)\nonumber \\
&=&\int d^3r_1 \int d^3 r_2 \,e^{-i{\mathbf
k}\cdot ({\mathbf r}_1-{\mathbf r}_2)} \left[ \rho_2 ({\mathbf r}_1,{\mathbf
r}_2)-n({\mathbf r}_1)n({\mathbf r}_2)\right]\;.
\label{s_i_k}
\end{eqnarray}
Here $\rho_2 ({\mathbf r}_1,{\mathbf r}_2)$ is the equal-times pair
distribution function, which for non-interacting fermions can be
expressed as 
\begin{equation}
\rho_{2}(\vett r_1,\vett r_2)=n(\vett r_1) \delta(\vett r_1-\vett
r_2)-\rho_{1}(\vett r_1,\vett r_2)\rho_{1}(\vett r_2,\vett
r_1)+n(\vett r_1) n(\vett r_2) \;,
\label{equal_times}
\end{equation}
where $\rho_{1}(\vett r_1,\vett r_2)=\langle\hat \psi^\dagger (\vett
r_1)\hat \psi(\vett r_2) \rangle$ is the one-body Dirac density matrix. 

The next sections will be devoted to the evaluation of these quantities
in a quasi-1D Fermi gas under harmonic confinement $V_{ext}(\vett r)=m
\omega_{ho}^2 [x^2+\lambda^2(y^2+z^2)]/2$, that is,
a Fermi gas in a cigar-shaped harmonic trap where $\lambda \gg 1$
and  only the values $n_y=0$ and $n_z=0$ of the  transverse
quantum numbers are allowed.
For simplicity we choose $\hbar=1$, $m=1$ and $\omega_{ho}=1$.

\section{Dynamic structure factor}
\label{secsw}
The transverse excited states
are  not involved in the excitation processes when {\it (i)} the
transverse component $k_\perp$ of the momentum transfer vanishes,
due to orthogonality of
harmonic oscillator wave functions, or {\it (ii)} the energy transfer $\omega$
is smaller than the gap between the chemical potential and the first
transverse excited state. 
In these limits Eq.~(\ref{s_i_kw}) reduces to a one-dimensional
problem and the 
transverse-state wave function factorizes out.
Following De Marco and Jin \cite{demarco_pra} we evaluate the overlap
integral $I=\int dx_1 \,\exp(i k_x x_1) \phi^*_i(x_1) \phi_j(x_1)$ in
momentum space to obtain
\begin{equation}
I=\int \frac{dq}{2 \pi} \,\tilde\phi^*_i(q+k_x/2) \tilde\phi_j(q-k_x/2)\;,
\end{equation}
where $\tilde \phi_i(p)={\cal H}_i(p)
\exp(-p^2/2)/\pi^{1/4}\sqrt{ 2^ii!}$.
By exploiting  the properties of the Hermite polynomials ${\cal H}_i(p)$
 \cite{gradshteyn}, we finally
obtain the expression for the dynamic structure factor as 
\begin{equation}
{\cal S}(\vett k, h)=2 \pi e^{-k_\perp^2/2\lambda} e^{-k_x^2/2}\sum_{i={\rm max}\{N-h,0\}}^{N-1}
\frac{i!}{(i+h)!} \left(\frac{k_x^2}{2}\right)^h 
\left[L_i^h\left(k_x^2/2 \right)\right]^2 \;.
\label{skw_exact}
\end{equation}
Here $h$ is an integer corresponding to a single-atom excitation
of $h$ quanta of the harmonic oscillator and  $L_i^h(x)$ is the $i^{th}$
generalized Laguerre
polynomial of parameter $h$. 
The spectrum in Eq.~(\ref{skw_exact}) is  discrete due to
the external harmonic confinement, but can of course be
approximated as a continuum in the limit $\omega \gg \omega_{ho}$. 
Figure~\ref{fig_sw} illustrates the spectrum at various numbers of
particles, for two values of the transferred wave vector in the $x$
direction.

The main qualitative features of the spectrum  can be understood 
within  a local-density description, which exploits  the results
for the dynamic structure factor $S_{hom}(k_x,\omega)$ 
of the homogeneous 1D system 
in a medium with a locally varying density
profile $n(x)$ to predict the spectrum  of a strictly
1D system as
\begin{equation}
S_{LDA}(k_x,\omega)= \int dx\, n(x) S_{hom}(k_x,\omega;\mu(x))\;. 
\label{slda}
\end{equation}
The spatial dependence of the chemical potential $\mu=k_F^2/2=\pi^2n^2/2$
is determined by the
relation $\mu(x)=\mu(n(x))$.
In the case of external harmonic confinement  we use the expression
for the homogeneous 1D Fermi gas, 
\begin{equation}
S_{hom}(k_x,\omega)= \left \{ \begin{array}{l l}
\pi/k_x k_F & \;\;\;{\rm if}\; |\omega_2(k_x)|<\omega<\omega_1(k_x) \\ & \\
0 & \;\;\;{\rm otherwise}\end{array}\right.
\end{equation}
 where
$\omega_{1,2}=k_x^2/2\pm k_x k_F$,
together with the density profile in LDA,
\begin{equation}
n(x)=\frac{1}{\pi}\sqrt{2N-x^2} 
\end{equation} 
to obtain the dynamic structure factor:
\begin{equation}
S_{LDA}(k_x,\omega)=\frac{2}{
k_x}\left[\sqrt{2N-\left(\frac{\omega}{k_x}-\frac{k_x}{2}\right)^2} 
-\sqrt{2N-\left(\frac{\omega}{k_x}+\frac{k_x}{2}\right)^2}\right]\;.
\label{skw_lda}
\end{equation}
The LDA curves for the dynamic structure factor 
are also shown in Fig.~\ref{fig_sw}, together with the spectrum of the
homogeneous  gas. Evidently, the effect of the confinement is
to change very substantially  
 the spectrum with respect to the homogeneous case, 
and the general spectral shape is well described within a
local-density picture. 

\section{Angular distribution of the scattered light}

\label{sk}
The angular distribution of the scattered radiation is related through
Eqs.~(\ref{s_kw}) and (\ref{essek}) to the 
equal-times pair distribution function.  
Both an elastic term and an inelastic term contribute to its expression:
the former is related to the
particle-density profile of the cloud while the latter arises from
the internal excitations of the cloud.

\subsection{Elastic contribution}
The elastic contribution to the diffraction pattern of  a quasi-1D  gas
is given by
\begin{equation}
S_e(\vett k) = e^{-k_\perp^2/2\lambda} \left|\int dx \, e^{-i k_x
x} n(x) \right|^2 \;.
\end{equation}
In evaluating this term, which is of order $N^2$, we have
looked  for the effect of the shell structure of the density profile
on the angular distribution of the scattered light.

With this aim, we have compared the Fourier Transform (FT) $\tilde
n(k_x)$ of the 
1D density profile with its LDA expression, which
can be evaluated  in terms
of the Bessel function $J_1$ as \cite{gleisberg}
\begin{equation}
\tilde{n}_{LDA}(k_x)=\dfrac{\sqrt{2N}}{k_x}J_1(\sqrt{2N}k_x)\;.
\end{equation}
The same behaviour is found for these two curves in the
central peak of the diffraction pattern
(Fig.\,\ref{fig0}(a)), where the dominant  contribution 
is due to the finite size of the cloud.
The FT of the exact profile differs instead from that of the LDA
(Thomas-Fermi) 
profile in that it contains  a side  peak at a value  $k_x\simeq
2\pi/\Delta  x$ of the transferred
momentum, 
where $\Delta x$ is the
average distance between two nearest maxima in the shell structure.
An estimate of $\Delta x$ can be made by taking  the maxima to be
uniformly distributed along the LDA profile, yielding 
\begin{equation} 
\Delta x\simeq\dfrac{2R_{LDA}}{N}=\dfrac{2\sqrt{2}\, a_{ho}}{\sqrt{N}},
\end{equation}
where $R_{LDA}$ is the axial Thomas Fermi radius of the cloud.
For $N$=100, $500$ and $1000$ we have recognized this side peak in the
diffraction pattern   (see Fig.\,\ref{fig0}(b) for $N$=100). As noticed
earlier  \cite{gleisberg}  
its position corresponds to $k_x\simeq 2 k_F$.
However, the relative intensity of the peak is extremely  weak and as we
expected it decreases 
with increasing $N$, as the density profile gets closer to the
Thomas-Fermi one.
At higher values of the momentum transfer the elastic contribution to
the diffraction pattern 
approximately vanishes, while the Thomas-Fermi pattern has an oscillating
tail which still reflects  the finite extension of the profile in
space.  

The total distribution of the scattered light is  the sum of
the elastic and inelastic contributions: while for small values of
$k_x$ the elastic contribution 
is dominant but does not distinguish between the true and the
Thomas-Fermi profile, the inelastic contribution
overcomes it at finite values of $k_x$. This is shown immediately
below.
 It is thus very difficult to
resolve the specific effect of the shell structure by this method. 

\subsection{Inelastic contribution: the static structure factor}

The inelastic contribution  
to the intensity of  scattered radiation is obtained from the
static structure factor in
Eq.~(\ref{s_i_k}).  This contribution is proportional to the number
$N$ of particles in the sample.
In the case of quasi-1D harmonic confinement, by the same procedure
which has led to Eq.~(\ref{skw_exact}) we obtain an analytic
expression for $S_i({\mathbf k})$
 in terms of the generalized Laguerre
polynomials: 
 \begin{equation}
S_i(\vett k)=N-e^{-k_\perp^2/2 \lambda}e^{-k_x^2/2} \sum_{i,j=0}^{N-1}
\frac{i!}{j!}\left(\frac{k_x^2}{2}\right)^{j-i}
\left[L_i^{j-i}(k_x^2/2)\right]^2\;.
\end{equation}
In Figure~\ref{fig_sk_exact} we show the function $S_i(\vett k)$ at
$k_\perp=0$ for 
various values of the particle number $N$. 
The condition $k_\perp\simeq 0$ applies to small 
angles in Fig.~\ref{fig_1}.
The quantum effect of single-level occupancy 
in the external potential is most visible in the first derivative
of $S_i(k_x)$, as is shown in the inset in
Fig.~\ref{fig_sk_exact}. 

The local-density result, obtained by integration of the 1D
Eq.~(\ref{skw_lda}) 
yields the following expression for a strictly 1D system:
\begin{equation}
S_{LDA}(k_x)=
\left\{\begin{array}{ll} \left[2N 
 {\rm ArcTan}(k_x/\sqrt{8N -k_x^2})+k_x\sqrt{8N-k_x^2}/4
 \right]/\pi & {\rm if}\; 
 k_x<2\sqrt{2N}  \\ &
\\ N & {\rm if} \;k_x \ge 2 \sqrt{2N}\;.
\end{array}\right.
\end{equation}
It follows from the discussion in Sec.~\ref{secsw} that the 1D LDA
provides a good description of the structure factor at $k_\perp=0$
as $N$ increases. 
In fact, although  the LDA  does not take
shell  effects into account, it  
gives the general behavior of $S_i(k_x)$  already for
small  numbers of fermions ($N\ge 20$). The LDA curve 
has not been shown in
Fig.~\ref{fig_sk_exact}  since it is indistinguishable from the
$N=100$ result.

\section{Equal-times pair distribution function}
\label{sec6}

Because of its definition as a double space integral in
Eq.~(\ref{s_i_k}), the inelastic term in the static structure factor
displays only part of the information which is contained in the
equal-times pair distribution function $\rho_{2}(\vett r_1,\vett r_2)$ as
defined in Eq.~(\ref{equal_times}). In this section we give a full
evaluation of this function for the quasi-1D gas of present
interest. Our main objective will be to explicitly show how
$\rho_{2}(\vett r_1,\vett r_2)$ reflects the Pauli exclusion principle
and the shell structure of the fermion cloud. The same results 
apply to  a 1D  Bose gas with hard-core interactions, 
since  the boson-fermion mapping holds  for the 
equal-times pair distribution function \cite{girardeau_p}.

With the notation $\phi_\perp(y,z)$ for the transverse ground-state
wave function of
the quasi-1D system of non-interacting fermions in the external
confining potential, we can write the pair distribution function as 
\begin{equation}
\rho_{2}(\vett r_1,\vett r_2)=n(\vett r_1)\delta(\vett r_1-\vett r_2)-
w(\vett r_{1\perp},\vett r_{2\perp})\left[F(x_1,x_2)-
n(x_1)n(x_2)\right]
\label{krunk}
\end{equation}
where $r_{\perp}=(y,z)$ and $w(\vett r_{1\perp},\vett r_{2\perp})=
|\phi_\perp(y_1,z_1)|^2 |\phi_\perp(y_2,z_2)|^2$. The 
function $F(x_1,x_2)$ is given by 
\begin{equation} 
F(x_1,x_2)=\sum_{i,j=1}^N  \phi_i^*(x_1) \phi_j(x_1)
\phi^*_j(x_2) \phi_i(x_2)
\label{effe}
\end{equation}
and will be calculated below by an efficient numerical method based on
an extension of the Green's function method used previously
\cite{noi,papirone1} to evaluate particle and kinetic energy
densities. 
 Direct calculation can be used for limited numbers of fermions
\cite{girardeau_p}, but the present method is easily applied  
to handle mesoscopic clouds containing a large number of fermions. 

\subsection{Method}

We first rewrite Eq.~(\ref{effe}) in terms of the Green's function
$\hat G(x)=(x-\hat x +i \varepsilon)^{-1}$ in
coordinate space, where
$\hat x$ is the position operator. We obtain 
\begin{eqnarray}
F(x_1,x_2)&=&\frac{1}{\pi^2} \lim_{\varepsilon \rightarrow 0^+}
\sum_{i,j=1}^N \Im \langle \phi_i | \hat G(x_1) |\phi_j
\rangle \, \Im \langle \phi_j | \hat G(x_2) |\phi_i
\rangle \nonumber \\ 
&=&\frac{1}{\pi^2} \lim_{\varepsilon \rightarrow 0^+} \Tr \left\{
\Im\left[ \hat G_N (x_1)\right] \cdot\Im\left[ \hat G_N (x_2)\right] \right\}
\;, 
\end{eqnarray}
where $|\phi_i \rangle$ are the eigenstates of the 1D Hamiltonian in the
coordinate representation and $\hat G_N (x)$ is the first $N\times N$
block of the matrix $\hat G(x)$. 
We then use the property $\Im A \cdot \Im B=(1/2)[\Re (A \cdot
B^*)-\Re (A \cdot B)]$ to obtain
\begin{equation}
F(x_1,x_2)=\frac{1}{2\pi^2} \lim_{\varepsilon \rightarrow 0^+} \left\{
\Re \Tr \left[ \hat G_N (x_1) \cdot  \hat G^*_N
(x_2)\right] - \Re \Tr \left[ \hat G_N (x_1)
\cdot \hat G_N (x_2)\right] \right\}\;.
\end{equation}
We can now use the relation $\Tr \,Q^{-1}=[\partial
\log\det (Q+\lambda {\Bbb I})/\partial \lambda]_{\lambda=0}$~\cite{fgv},
 with ${\Bbb I}$
being the identity matrix $N\times N$, to obtain
the final expression
\begin{eqnarray}
F(x_1,x_2)&=&\frac{1}{2\pi^2} \lim_{\varepsilon \rightarrow 0^+} 
\Re \frac{\partial}{\partial \lambda}\left[\log \det \left((x_2-\hat
\xi^*(x_2)-i\varepsilon)\cdot(x_1-\hat \xi(x_1)+i\varepsilon)+ \lambda
{\Bbb I}\right)\right]_{\lambda=0}\nonumber \\ 
&-& \frac{1}{2\pi^2} \lim_{\varepsilon \rightarrow 0^+} 
\Re \frac{\partial}{\partial \lambda}\left[\log \det \left((x_2-\hat
\xi(x_2)+i\varepsilon)\cdot(x_1-\hat \xi(x_1)+i\varepsilon)+ \lambda
{\Bbb I}\right)\right]_{\lambda=0}\;.
\label{log_det}
\end{eqnarray}
Here, $\hat \xi(x)$ is a renormalized position operator \cite{cipro}, 
reduced to  the $N\times N$ subspace.

While Eq.~(\ref{log_det}) holds for any 1D  confining
potential, this approach is particularly useful for the case of
harmonic confinement
where the representation of the position operator
is tridiagonal. In this case the renormalized operator $\hat \xi (x)$
reads  $[\hat \xi(x)]_{i,j}=[\hat x]_{i,j}$ if $(i,j)\neq(N,N)$ 
and  $[\hat \xi(x)]_{N,N}=\tilde x_{N,N}(x)$, with
\begin{equation}
\tilde{x}_{N,N}(x)=\cfrac{N/2}{x+i\varepsilon-\cfrac{(N+1)/2}
{x+i\varepsilon-\dots}}\;.
\label{contfr}
\end{equation}
The evaluation of $F(x_1,x_2)$ is thus reduced to the
calculation of the determinant of pentadiagonal  $N\times N$ matrices.
If we make a partition into blocks $A_i$ and $B_{i,j}$ of dimension
$2\times2$, a pentadiagonal matrix $Q$ takes a tridiagonal form.
For even $N$ we can write (see also \cite{papirone1})
\begin{equation}
Q=\left(
\begin{array}{cccc}
A_1&B_{1,2}&&\\
B_{2,1}&A_2&B_{2,3}&\\
&B_{3,2}&\ddots&\ddots\\
&&\ddots&A_{N/2}\\
\end{array}\right)\;,
\end{equation}
and the determinant of such a matrix can be factorizes into the product
of $2\times2$ matrices as
\begin{equation}
\det Q=\prod_i\det{\tilde A_i}\;.
\end{equation}
Here,  $\tilde A_1=A_1$ and $\tilde
A_i=A_i-B_{i,i-1}(\tilde A_{i-1})^{-1}B_{i-1,i}$ for $i>1$.
If $N$ is odd we obtain an analogous expression in which  the last
block  of the partition is a  $1\times 1$ matrix.

\subsection{Numerical results}
By the method outlined above we have calculated the longitudinal
contribution to the equal-times pair distribution function, 
$\rho^{(l)}_2(x_1,x_2)=n(x_1)n(x_2)-F(x_1,x_2)$, for up to $N=100$
fermions without particular numerical
efforts.
Since this function  presents in general a number of maxima of 
order  $N^2$, for the sake of clarity we have shown  in
Fig.~\ref{fig3D}  the full result only for the case of $N=4$ fermions. 

The numerical results  for higher values of the particle number $N$ are 
shown in Fig.~\ref{figseztot} through a  section of the pair
distribution function  taken at the
value  $R\equiv(x_1+x_2)/2=0$ of the center-of-mass coordinate. In the
confined gas  the effect
of Pauli
exclusion  is also observable in real space as a depression of
$\rho_{2}^{(l)}(x_1,x_2)$ at short distance $r\equiv(x_1-x_2)$.
This  is more clearly
illustrated in the inset of Fig.~\ref{figseztot}  through an
enlargement of the region near $r=0$.
At larger distances the function $F(x_1,x_2)$ goes rapidly
to zero and the pair function decouples into the product of two
particle  density profiles.    
For other sections of the pair function a similar behavior is found
with a reduced number of peaks in the profile.
This is shown in
Fig.~\ref{figsezN4}  for the illustrative case of $N=4$ fermions.

\section{Summary and conclusions}
\label{sec7}

In summary, we have evaluated the scattering functions of quasi-1D
mesoscopic clouds of non-interacting fermions in an external harmonic
potential. A number of conclusions can be drawn from our results.

\begin{figure}
\centerline{\psfig{figure=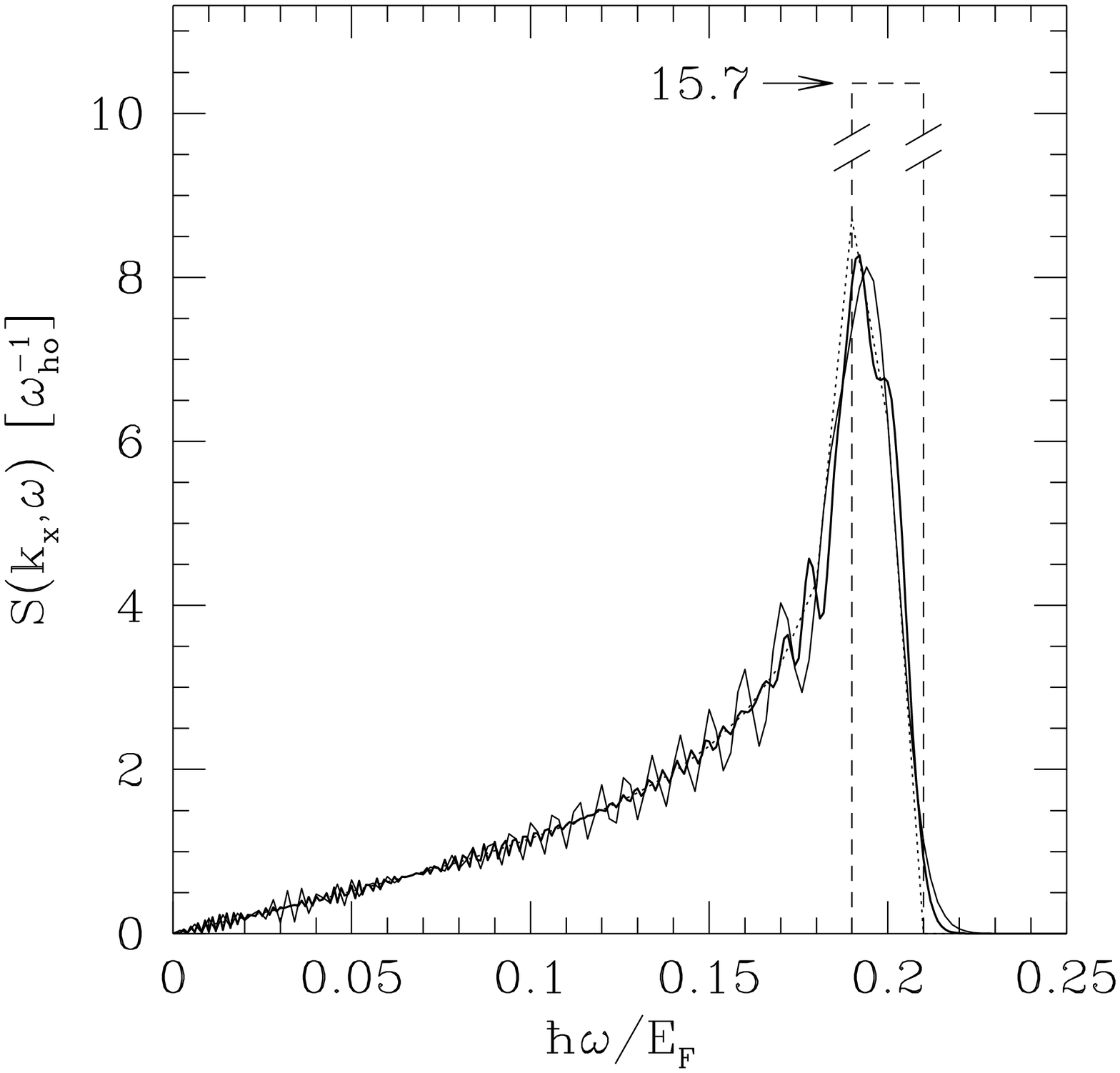,width=0.47\linewidth}\psfig{figure=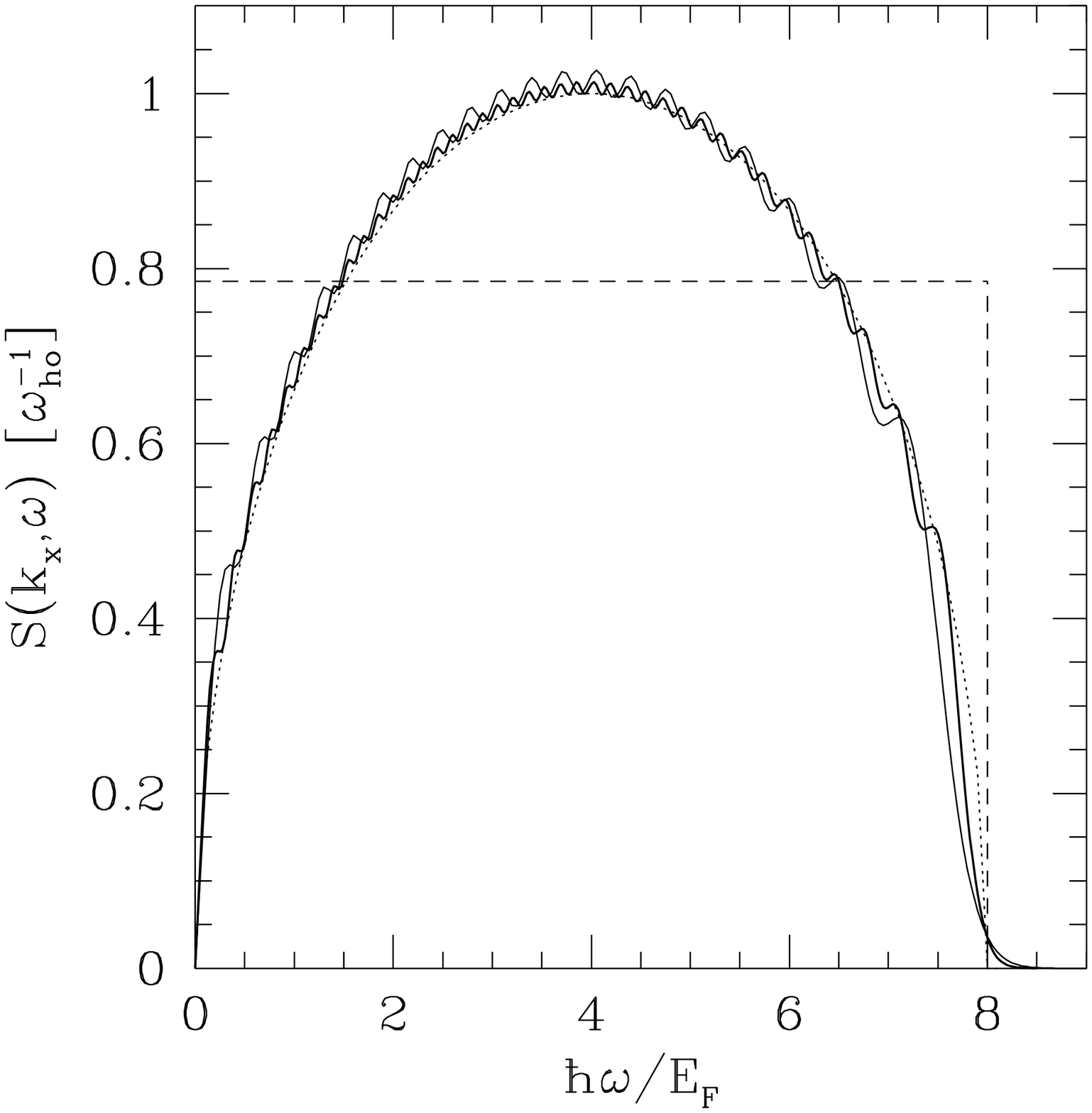,width=0.47\linewidth}}   
\caption{Dynamic structure factor $S(k_x,\omega)$ 
of a  1D harmonically confined Fermi gas (in units of
$\omega_{ho}^{-1}$) 
as a function of  $\hbar \omega/E_F$,
where $E_F=N \hbar \omega_{ho}$.
Left panel: at  $k_x=0.1$ $k_F$ and
numbers of fermions $N$=500 (solid line) and 
$N$=1000 (solid bold line).
 Right panel: at  $k_x=2$ $k_F$ and  $N$=20 (solid
line) and  $N$=40
(solid bold line). 
In both panels 
the LDA spectrum (dotted line) and the  dynamic structure factor of 
a homogeneous 1D Fermi gas (dashed line) are plotted  in the same units.
The exact $S(k_x,\omega)$ is shown for clarity as
constructed from discrete frequencies by
joining vertical lines having height proportional to the oscillator
strength.}
\label{fig_sw}
\end{figure}

The inelastic scattering spectrum
of the inhomogeneous  gas, displayed in
Fig.~\ref{fig_sw},  reflects its restricted
geometry through {\it (i)} the form factor 
coming from the tight  transverse confinement, and {\it (ii)}  its
general shape,
which is remarkably different from that  of the  homogeneous   
gas, as can be predicted already in a local-density calculation 
(see Eqs.~(\ref{slda})-(\ref{skw_lda})). 
In principle the spectral intensity has a discrete
structure owing to the level spacings induced by the axial
confinement, but this spectral structure is rapidly smoothed away 
as the number $N$ of fermions in the cloud increases.

\begin{figure}
\centerline{\psfig{figure=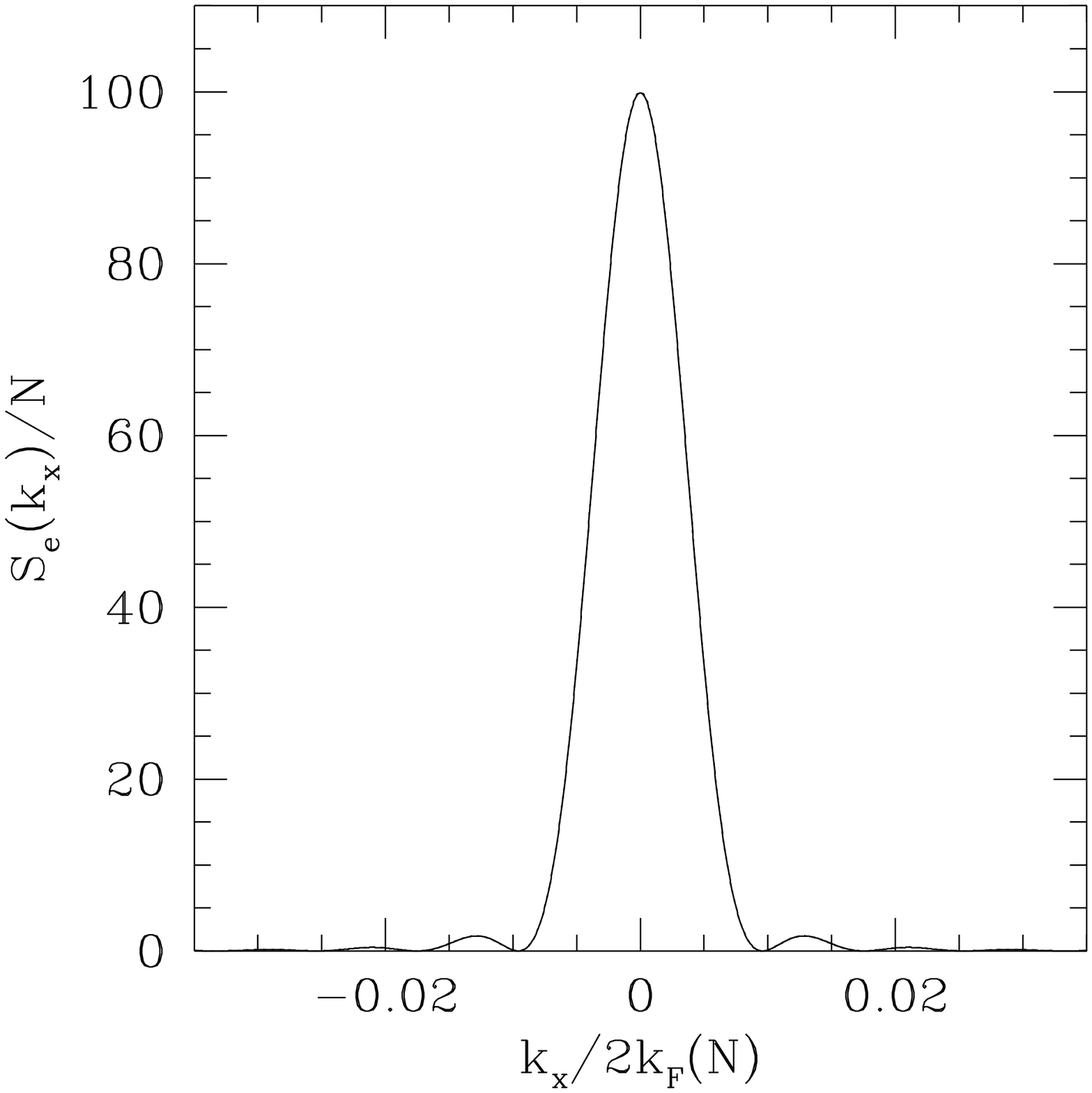,width=0.47\linewidth}\psfig{figure=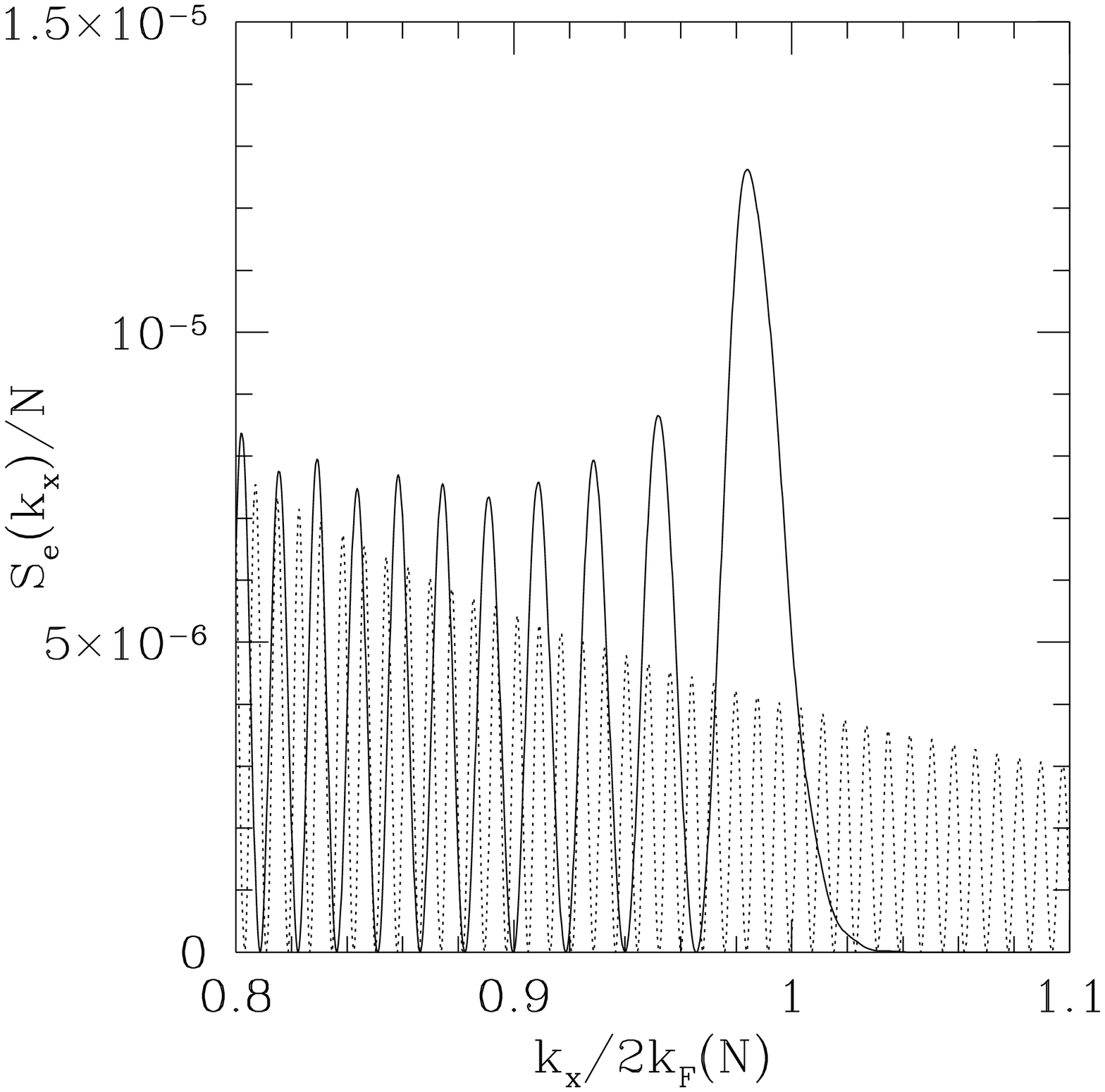,width=0.47\linewidth}}
\caption{Central peak (left panel) and  side peak (right panel)
of the elastic contribution to the static structure factor, normalized
to $N=100$ fermions, as a function of the
wave number  $k_x$ (in units of $2k_F(N)=2\sqrt{2N}$ $ a_{ho}^{-1}$)  at fixed 
$k_\perp=0$. Both  the
true profile  (continuous line)
and the LDA  
profile (dotted line) are shown.}
\label{fig0}
\end{figure}

The elastic contribution to the scattering cross section is dominated
by the main peak centered at zero momentum transfer and broadened as a
consequence of the finite axial size of the cloud (see Fig.~\ref{fig0}
(a)). The shell structure of the cloud appears in the elastic
scattering intensity only as an extremely weak peak located at
momentum transfer close to twice the Fermi momentum. This peak  lies in
the far tail of the central elastic peak (see Fig.~\ref{fig0} (b)) and
will be masked by the inelastic scattering contribution to the
diffraction pattern.

\begin{figure}
\centerline{\psfig{figure=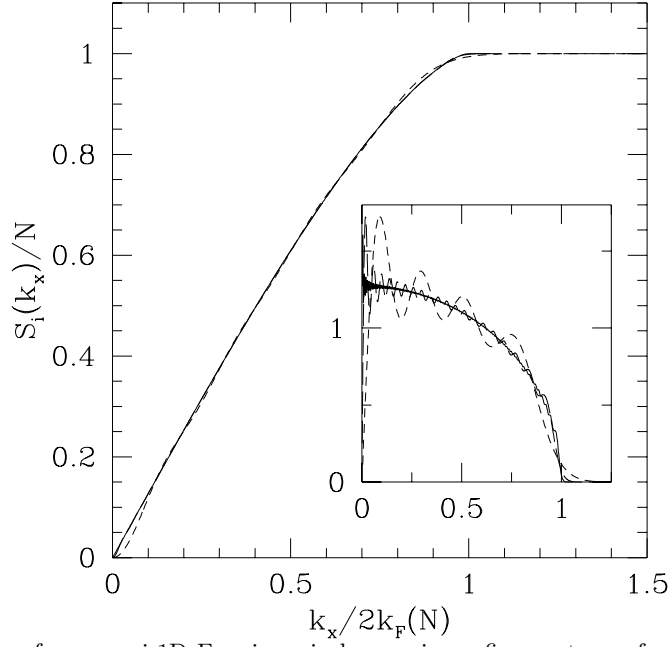,width=0.5\linewidth}}
\caption{Static structure factor for a quasi-1D Fermi gas in 
harmonic confinement as a function of the wave number  $k_x$  (in
units of $2k_F(N)$) at 
$k_\perp=0$. The inset shows the first
derivative of $S_i(k_x)$ in the same reduced units. The
numbers of fermions are
$N=$4 (short-dashed line), $N=$20 (long-dashed line) and  $N=$100 (solid
line). }
\label{fig_sk_exact}
\end{figure}

The inelastic contribution to the static structure factor in
Fig.~\ref{fig_sk_exact} is obtained by integration of the dynamic
structure factor over energy transfers. Similar comments apply,
therefore, to these two scattering functions: thus, $S_i(k_x)$
is  clearly reflecting the shell structure of the cloud only
for limited values of the number of fermions (see the inset of
Fig.~\ref{fig_sk_exact}). The general shape of $S_i(k_x)$
in the main body of Fig.~\ref{fig_sk_exact} is again related through
the local-density scheme to the static structure factor of the 1D
Fermi gas. 

\begin{figure}
\centerline{\psfig{file=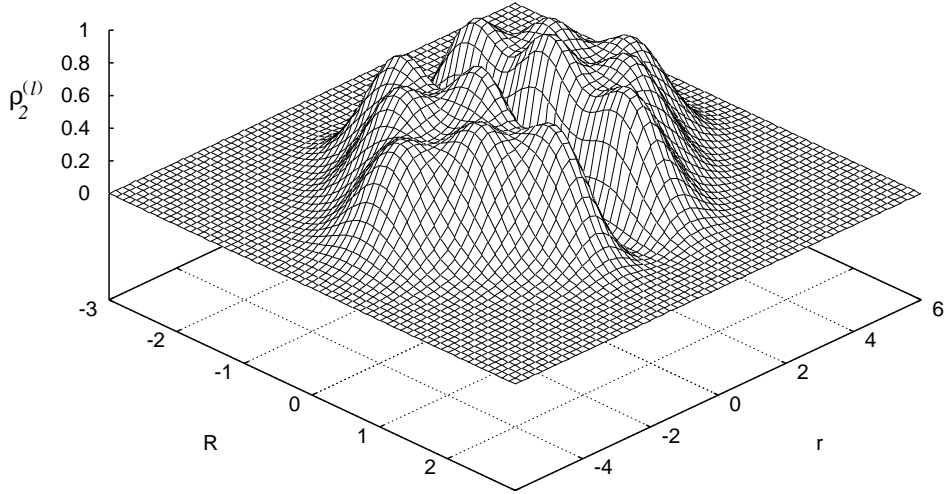,width=0.8\linewidth}}
\caption{Longitudinal contribution $\rho_{2}^{(l)}(r;R)$ 
to the pair distribution function  for $N$=4 fermions 
in quasi-1D  harmonic confinement,  in
units of $a_{ho}^{-2}$,
as a function of the
center-of-mass coordinate $R$  and of the relative
coordinate $r$, both in units of $a_{ho}$. }
\label{fig3D}
\end{figure}

\begin{figure}
\centerline{\psfig{file=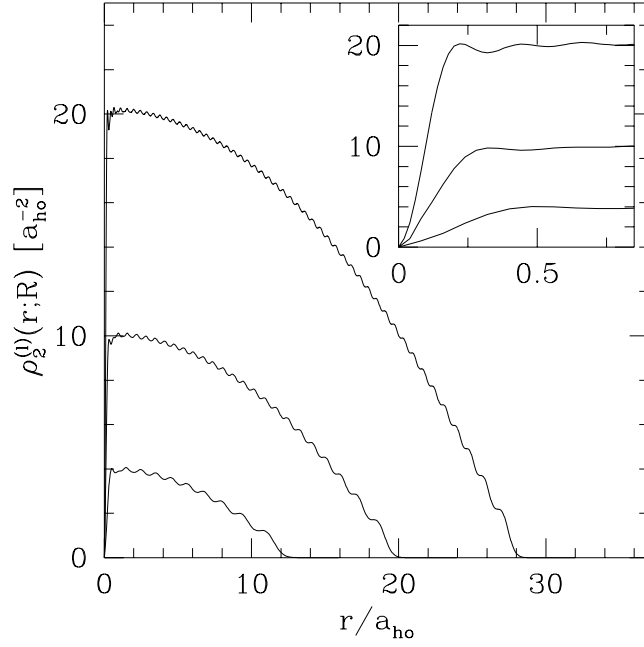,width=0.5\linewidth}}
\caption{Section of 
$\rho_{2}^{(l)}(r;R)$ at $R=0$
for $N$=20, 50 and 100 fermions at $R=0$ (from bottom to top), in
units of $a_{ho}^{-2}$, as a function of the
relative coordinate $r/a_{ho}$ of the pair.
 The inset shows an enlargement of the 
 region near $r=0$, in the same units.}
\label{figseztot}
\end{figure}

\begin{figure}
\centerline{\psfig{file=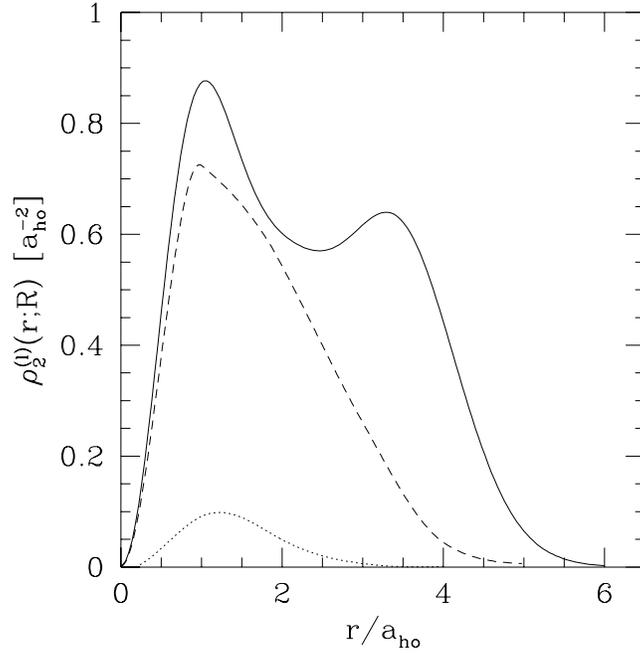,width=0.5\linewidth}}
\caption{Sections of $\rho_{2}^{(l)}(r;R)$  for values of
$R$=0 (solid line), $R=a_{ho}$ (dashed line) and $R=2 a_{ho}$ (dotted line) at
$N$=4. Notations and
units are  as in Fig.~\ref{figseztot}.}
\label{figsezN4}
\end{figure}

The equal-times pair distribution function 
contains a much greater and detailed
amount of information on the relative spatial distribution of pairs of
particles in the inhomogeneous Fermi gas. Here it is, of course, a
function both of the relative distance $r$ of the pair and of the
coordinate $R$ of its center-of-mass along the $x$ axis. The Pauli 
``exchange'' hole surrounding each fermion is a well-known feature of
the pair distribution function in homogeneous Fermi gases and is a
very prominent feature in the spin-polarized inhomogeneous system. It
is seen in Fig.~\ref{fig3D} as the deep valley running in a direction
parallel to the $R$ axis and cutting into two separate parts the contour
plot at $r=0$. It is again seen in the inset of Fig.~\ref{figseztot}
at $R=0$ and in Fig.~\ref{figsezN4} at various values of 
$R$. In addition to this basic feature, the two-body distribution
displays a great deal of secondary structures, of order $N^2$ as
already noted, which descend from the combination of the Pauli principle
and of the axial confinement.

In conclusion, the single occupancy of the discrete quantum levels in
the quasi-1D Fermi gas may become accessible in light scattering
experiments by looking at the inelastic scattering spectrum. It also
affects to some extent the angular distribution of the inelastically
scattered light. The elastic contribution to the scattering is quite
negligible in this respect.

\acknowledgements
A.M. acknowledges useful discussions with Dr. I. Carusotto.
This work was partially supported by MURST under the PRIN2000
Initiative.

\newpage

\end{document}